\documentclass[sigconf]{acmart}
\AtBeginDocument{%
  }

\copyrightyear{2026}
\acmYear{2026}
\setcopyright{cc}
\setcctype{by}
\acmConference[CIKM '26]{Proceedings of the 35th ACM International Conference on Information and Knowledge Management}{November 07--11, 2026}{Rome, Italy}
\acmBooktitle{Proceedings of the 35th ACM International Conference on Information and Knowledge Management (CIKM '26), November 07--11, 2026, Rome, Italy}
\acmDOI{10.1145/3799682.3840167}
\acmISBN{979-8-4007-2539-5/2026/11}

\usepackage{enumitem}
\usepackage{multirow}
\usepackage{balance}

\begin{document}

\title{Correcting to Predict: Pseudo-Value Correction for Multimodal Attribute Value Extraction}

\author{Junhao Zhang}
\authornote{Both authors contributed equally to this research.}
\authornote{Corresponding author}
\email{junhao.zjh@alibaba-inc.com}
\orcid{1234-5678-9012}
\author{Feiran Hu}
\authornotemark[1]
\email{hufeiran.hfr@alibaba-inc.com}
\affiliation{%
  \institution{Alibaba International Digital Commerce Group}
  \city{Hangzhou}
  \country{China}
}

\author{Xiao Hu}
\email{hx319433@alibaba-inc.com}
\affiliation{%
  \institution{Alibaba International Digital Commerce Group}
  \city{Hangzhou}
  \country{China}}

\author{Baoliang Cui}
\email{moqing.cbl@taobao.com}
\affiliation{%
  \institution{Alibaba International Digital Commerce Group}
  \city{Hangzhou}
  \country{China}
}

\author{Xiaoyi Zeng}
\email{yuanhan@taobao.com}
\affiliation{%
  \institution{Alibaba International Digital Commerce Group}
  \city{Hangzhou}
  \country{China}}

\renewcommand{\shortauthors}{Junhao Zhang, Feiran Hu, Xiao Hu, Baoliang Cui, and Xiaoyi Zeng}

\begin{abstract}
Product attribute value extraction (AVE) is a fundamental task in e-commerce, aiming to identify specific values of predefined attributes from multimodal product profiles such as text and images. While multimodal large language models (MLLMs) have shown promise for AVE, they face challenges in extracting implicit attributes that require joint reasoning over visual and textual cues, often confusing semantically similar values. However, existing methods often fail to resolve such ambiguities because the correct value often depends on subtle multimodal cues that are easy to miss or override. To address this challenge, we propose Correcting to Predict (C2P), a framework that treats attribute extraction as a correction process. Given an initial pseudo-value such as a retrieved candidate or placeholder, the model learns to correct it using multimodal evidence. During training, diverse pseudo-values help the model learn evidence-based correction behavior, and a self-consistency refinement stage further reduces sensitivity to pseudo-value perturbations. At inference, a fixed placeholder triggers the learned correction behavior, enabling efficient single-pass prediction without online retrieval or iterative refinement.
We evaluate C2P on a public benchmark and a large-scale industrial dataset. Offline results show that C2P outperforms strong baselines, with notable gains on ambiguous attributes. Online A/B tests on AliExpress further show consistent improvements in seller adoption, attribute completeness, and user engagement, validating C2P’s effectiveness and efficiency in real-world deployment.

\end{abstract}



\maketitle

\section{Introduction}
Product attributes are fundamental for e-commerce, enabling user navigation (e.g., faceted filtering)~\cite{more2016attribute, shinzato2013unsupervised}, powering algorithmic systems (e.g., semantic retrieval, recommendation, and question answering)~\cite{karamanolakis2020txtract}, and supporting business operations (e.g., demand forecasting, assortment optimization)~\cite{ghani2006text}.

Despite their importance, attribute values are often missing or misaligned due to unstructured seller input and optional completion~\cite{dong2020autoknow, more2016attribute}. As a result, only a small fraction of products have complete and reliable attribute annotations, severely limiting both user experience and algorithmic performance~\cite{zheng2018opentag, wang2020learning}. 
This necessitates automated attribute value extraction (AVE): identifying correct values from a product's title, description, and image, for predefined attributes without manual labeling.

Over the past decade, AVE has evolved significantly. Early named entity recognition (NER) approaches extracted explicit spans from text~\cite{more2016attribute, putthividhya2011bootstrapped}, but struggled when values were not verbatim in descriptions. Question answering frameworks improved generalization by treating attributes as queries~\cite{wang2020learning}, while hierarchical multi-task learning scaled to thousands of categories~\cite{karamanolakis2020txtract}. However, text-only methods faced fundamental limits in disambiguating attributes like whether ``golden'' refers to color or material. This spurred multimodal frameworks that jointly model textual and visual features~\cite{zhu2020multimodal}, yet challenges persisted for implicit AVE—inferring values not explicitly mentioned~\cite{zhang2023pay, zou2024eiven}.

However, applying multimodal LLMs (MLLMs) to implicit AVE remains challenging: benchmarks on \textit{ImplicitAVE}~\cite{zou2024implicitave} show significant performance drops on context-dependent attributes like \textit{Boot Style} or \textit{Season}~\cite{zou2024eiven}. \textbf{A key challenge is ensuring that predictions are adequately supported by multimodal evidence}, rather than by surface-level features or linguistic priors. 

In response, recent work has explored various strategies to improve reliability. For example, model ensembles combine predictions from multiple LLMs to improve accuracy~\cite{fang2024llm}. Multi-agent debate frameworks iteratively refine responses through cross-model critique~\cite{huang2025madiave}. Others explore self-correction mechanisms that prompt a model to revise its own output~\cite{brinkmann2025self}. While these approaches aim to improve robustness, they share a critical limitation: \textbf{their initial prediction step may still be insufficiently grounded in multimodal evidence}. Subsequent steps, such as voting, debate, or revision, attempt to refine this initial prediction, but early errors can persist or even be reinforced.

\begin{figure}[h]
  \centering
  \includegraphics[width=\linewidth]{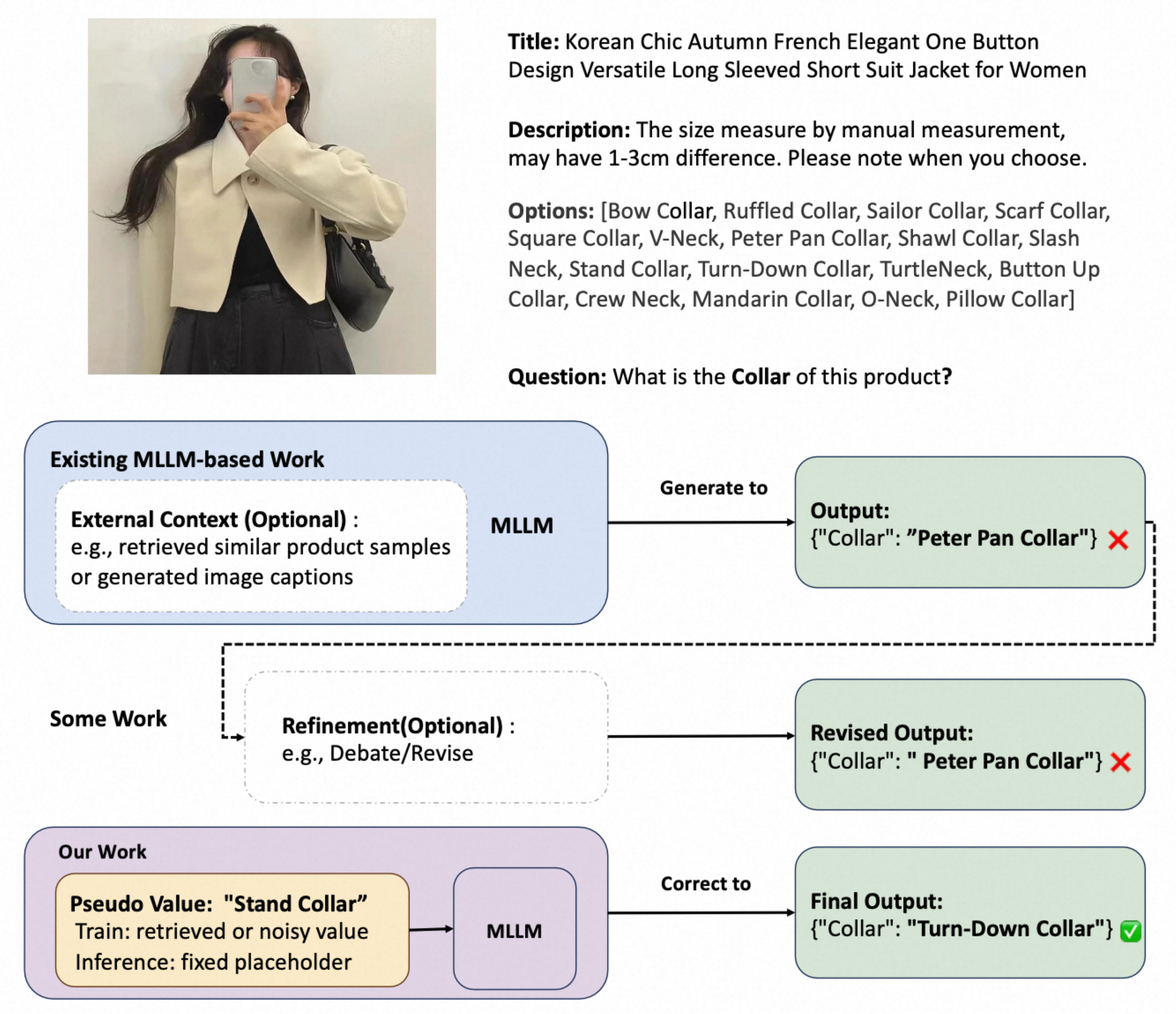}
  \caption{At inference, our C2P framework replaces external context with pseudo-value correction, enabling single-pass, retrieval-free prediction.}
\label{fig:framwork_comp}
\end{figure}

To address this challenge, we propose \textbf{Correcting to Predict (C2P)}, a framework that reformulates AVE as correcting an initial
pseudo-value rather than directly generating an attribute value.
The key idea is to train prediction under a supplied hypothesis:
the pseudo-value can be a retrieved candidate or a simple placeholder, and the model learns when it should be retained or corrected
according to multimodal evidence. Retrieved candidates, when available, are therefore treated as hypotheses to be verified rather than as trusted context to be copied. To further reduce sensitivity to pseudo-value perturbations, we introduce a self-consistency refinement (SCR) stage that identifies samples with inconsistent predictions under different pseudo-values and refines the model on these cases. At inference time, a fixed placeholder is sufficient to trigger the learned correction behavior, enabling efficient single-pass prediction without online retrieval or iterative refinement.

Our main contributions can be summarized as follows:

(1) We reformulate multimodal AVE as correcting an initial pseudo-value with multimodal evidence, providing an alternative to direct generation for implicit attribute extraction.

(2) We propose \textbf{Correcting to Predict (C2P)}, a framework that learns evidence-based correction behavior from diverse pseudo-
values, together with a Self-Consistency Refinement (SCR) stage that further reduces sensitivity to pseudo-value perturbations.

(3) We show that the learned correction behavior enables a simple deployment design based on a fixed placeholder, supporting retrieval-free single-pass inference with strong efficiency in large-scale production systems.

(4) We conduct extensive evaluations on the public \textit{ImplicitAVE} benchmark, a large-scale industrial dataset, and online A/B tests on AliExpress. The results show that C2P consistently outperforms strong baselines, especially on ambiguous attributes.

\section{Related Work}
\label{sec:related}

\paragraph{Direct Generation Approaches for AVE}
Recent AVE methods increasingly leverage large language models (LLMs) and multimodal LLMs (MLLMs) for end-to-end generation given a predefined attribute. ExtractGPT~\cite{brinkmann2024extractgpt} demonstrates strong zero-shot performance using GPT-4 and Llama-3, while EIVEN~\cite{zou2024eiven} pioneers MLLMs for implicit attribute value extraction, using multi-granularity visual features and a Learning-by-Comparison (LBC) technique to reduce confusion among similar values. MICE~\cite{gong2025mice} further enhances inputs with offline-generated captions from multiple MLLMs. Despite their effectiveness, these approaches treat AVE as direct generation from multimodal context, without explicitly training the model to verify or revise its outputs against multimodal evidence.

\paragraph{Retrieval-Augmented Generation for AVE}
MVP-RAG~\cite{zou2025multi} addresses the broader Product Attribute Value Identification (PAVI) task by retrieving similar products and their attribute values to condition a generative model. Although designed for PAVI, it can be adapted to AVE by fixing the target attribute $a$. In this setting, retrieved candidates serve as reference context. In contrast, C2P treats retrieved values as hypotheses to be verified and corrected against multimodal evidence. As a result, MVP-RAG does not explicitly train the model to detect and correct mismatches between retrieved references and the current product's visual-textual cues, which can propagate retrieval errors and increase inference latency.

\paragraph{Post-Generation Refinement in AVE}
Self-correction methods first generate an initial, potentially incorrect output, then prompt the model to revise it~\cite{brinkmann2025self}. However, empirical studies show limited gains: self-correction often fails to improve accuracy despite higher computational cost~\cite{brinkmann2025self}. Multi-agent debate (e.g., MADIAVE~\cite{huang2025madiave}) improves robustness by iterative refinement through cross-model critique. While these strategies enhance reliability, they share a fundamental constraint: correction occurs \textit{after} the initial prediction, which may already be biased due to insufficient grounding in multimodal evidence. Consequently, these post-hoc mechanisms often struggle to effectively override deeply entrenched initial errors.

\paragraph{Connection to Curriculum Learning}
Our Self-Consistency Refinement (SCR) is inspired by curriculum learning (CL)~\cite{bengio2009curriculum}, which improves robustness by progressively introducing challenging samples. Conventional CL typically defines sample difficulty using generic signals such as training loss or prediction confidence~\cite{wang2021survey}. In contrast, SCR uses a task-specific signal: prediction inconsistency under pseudo-value perturbation. This helps focus training on cases where the model’s output changes when the input value is altered, indicating that the prediction may rely too much on the input guess rather than the actual product evidence. Unlike generic curriculum learning, SCR directly targets the failure mode of interest, namely prediction instability under pseudo-value perturbation.

\paragraph{Synthesis and Our Position}
Crucially, C2P embeds verification into prediction by treating pseudo-values, including retrieved candidates, as \textit{correctable hypotheses} rather than contextual references (as in RAG) or post-hoc refinements. At inference, it uses a fixed placeholder instead of retrieval, enabling retrieval-free, single-pass prediction based on the learned correction behavior.

\section{Problem Statement}
\label{sec:problem_statement}

We consider the task of \textbf{Multimodal Product Attribute Value Extraction (AVE)} in e-commerce. Formally, given a product profile with an image $I$, textual fields $T=(T_t, T_d)$ consisting of a title $T_t$ and an optional description $T_d$, and a predefined attribute name $a$, the task is to predict the correct value $v$ for attribute $a$ using evidence from both $I$ and $T$.

The extracted value \(v\) should satisfy two key requirements:
\begin{enumerate}[leftmargin=*]
    \item \textbf{Evidence-based}: $v$ must be justified by visual and textual evidence. If the attribute is inapplicable to the product (e.g., \textit{Shoe Size} for a dress) or no sufficient evidence exists, the model outputs \texttt{invalid}.
    \item \textbf{Canonical}: If $v \neq \texttt{invalid}$, it must be selected from the platform's Canonical Product Vocabulary (CPV) for $a$ (e.g., \texttt{crew neck} not \texttt{round collar}). This ensures that all model outputs are standardized, enabling reliable exact string matching during consistency checks and downstream processing.
\end{enumerate}

Let $\mathcal{O}_{c,a} = \mathcal{V}_{c,a} \cup \{\texttt{invalid}\}$ denote the output space for attribute $a$ in category $c$, where $\mathcal{V}_{c,a}$ is the finite set of valid canonical labels defined in the platform's Canonical Product Vocabulary (CPV) for the (category, attribute) pair $(c, a)$. Each label in $\mathcal{V}_{c,a}$ is a single standardized string, which may represent a composite concept (e.g., \texttt{Cotton,Polyester}). The AVE task is thus formalized as mapping $(I, T, c, a)$ to a value $v \in \mathcal{O}_{c,a}$.

\textbf{Key Challenge}: This task is especially difficult for \textit{implicit attributes}—values not explicitly mentioned in text and requiring joint visual-textual reasoning (e.g., inferring \texttt{turtleneck} from an image of a high-collar sweater when the title states only \texttt{winter top}). Real-world profiles often contain noisy, missing, or conflicting cues, demanding models that actively justify predictions against multimodal evidence rather than relying on surface-level patterns.

\begin{figure*}[t]
  \centering
  \includegraphics[width=\linewidth]{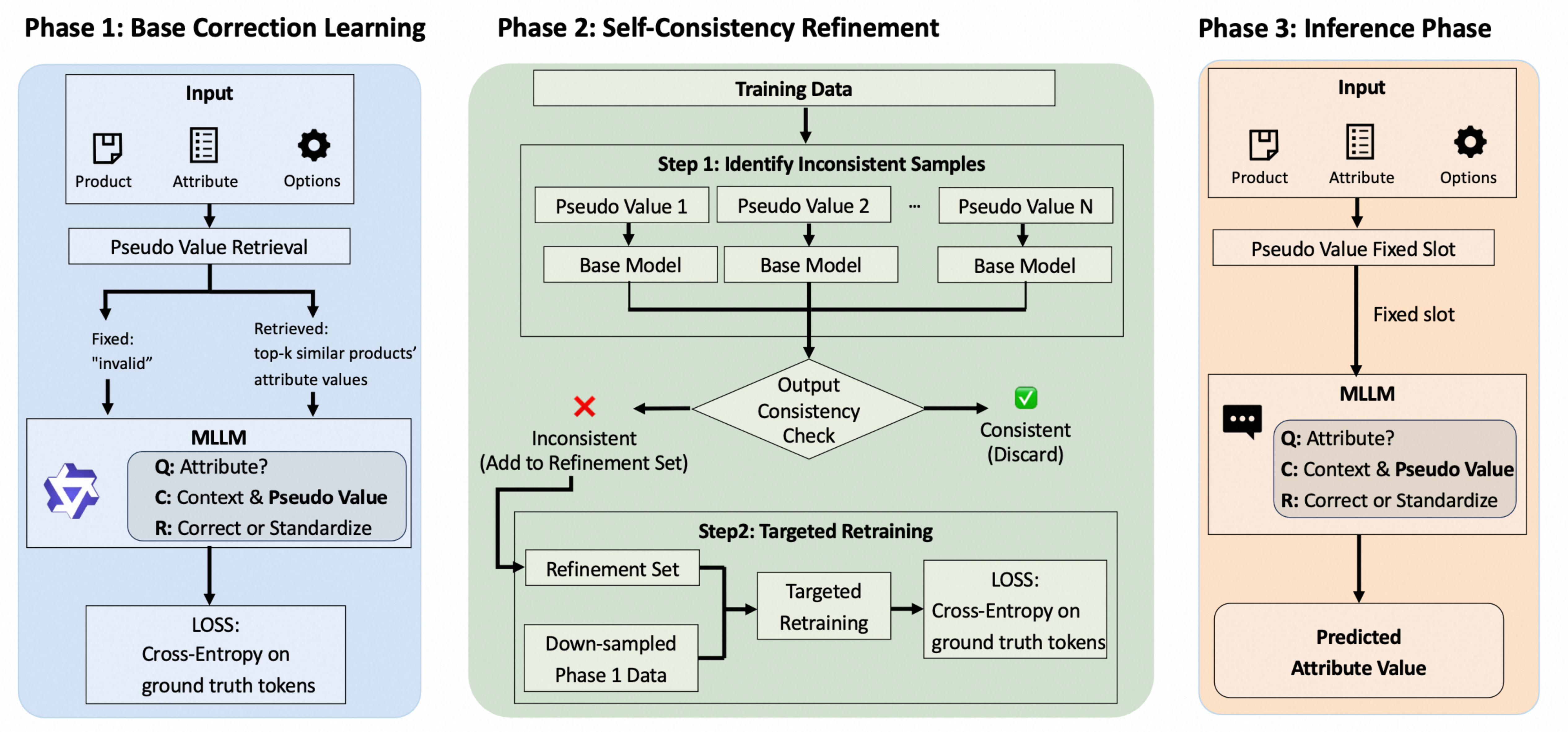}
  \caption{
  Overview of the C2P framework. 
(1) Base training: the model learns to correct diverse pseudo-values using multimodal evidence.
(2) Self-Consistency Refinement (SCR): inconsistent samples are retrained to reduce pseudo-value sensitivity and encourage greater reliance on product evidence. 
(3) Inference: a fixed placeholder enables retrieval-free, single-pass prediction.
}
  \label{fig:framework}
\end{figure*}

\section{Methods}
\label{sec:methods}

We present \textbf{Correcting to Predict (C2P)}, a framework that reformulates multimodal AVE as correcting an initial pseudo-value with multimodal evidence. We first describe the basic C2P formulation and the construction of train-time pseudo-values. We then introduce a Self-Consistency Refinement (SCR) stage to further reduce sensitivity to pseudo-value perturbations. Finally, we describe the inference design of C2P, which uses a fixed placeholder for single-pass prediction without online retrieval or iterative refinement.

\subsection{C2P as Correction with Pseudo-Values}
\label{subsec:c2p}

Given a product with image $I$, title $T_t$, optional description $T_d$, category $c$, and target attribute $a$, the goal of AVE is to predict the canonical attribute value $y$. Conventional approaches formulate this task as direct generation from the multimodal product profile. In contrast, C2P introduces an explicit pseudo-value $v_{\text{pseudo}}$ and reformulates AVE as correcting this initial hypothesis according to multimodal evidence.

Accordingly, the model is trained on structured prompts that explicitly encode the correction task. For a sample with pseudo-value $v_{\text{pseudo}}$ and a prompt option subset $\mathcal{C}_{c,a}^{(\leq 20)} \subseteq \mathcal{V}_{c,a}$, the input prompt is:

\begin{quote}
\small
\texttt{Question: What is \{a\} of this product?} \\
\texttt{Context: [Category] \{c\} <img> [Title] \{T\_t\}} \\
\texttt{~~~~~~~~[Description] \{T\_d\}} \\
\texttt{[Pseudo-Value] \{v\_pseudo\}} \\
\texttt{[Options] }[$o_1$, $o_2$, $\ldots$, $o_{k}$] (up to 20 most frequent values from $\mathcal{V}_{c,a}$) \\
\texttt{[Rule] If image or title or description contradicts the pseudo-value, correct it to the accurate value; otherwise, keep it but use the standard expression if available in options.} \\
\texttt{Answer:}
\end{quote}

The option set $\mathcal{C}_{c,a}^{(\leq 20)}$ contains up to 20 frequent canonical values for attribute $a$ under category $c$, derived from the platform's Canonical Product Vocabulary (CPV). Although only a subset of options is shown in the prompt, the model can still output any value in $\mathcal{O}_{c,a}$ when supported by multimodal evidence. The options therefore serve as soft guidance rather than hard constraints.

Training uses standard supervised fine-tuning. Let $y = (y_1, \ldots, y_L)$ denote the tokenized ground-truth canonical value. The base training loss is:
\begin{equation}
    \label{eq:base_loss}
    \mathcal{L}_{\text{base}} = -\sum_{t=1}^{L} \log P(y_t \mid I, T, c, a, v_{\text{pseudo}}, y_{<t}; \theta),
\end{equation}
where $\theta$ denotes the trainable parameters.

\subsection{Train-time Pseudo-Values}
\label{subsec:pseudo_values}

A key design choice in C2P is how to construct train-time pseudo-
values so that the model learns when an initial hypothesis should
be retained or corrected. We use two complementary types of pseudo-values.

\textbf{(1) Placeholder pseudo-values.}
We use simple placeholders such as \texttt{invalid}, which carry little useful information for the current sample and encourage the model to rely more heavily on multimodal evidence. However, such placeholders can also be relatively easy to bypass, especially on datasets where most samples have valid answers.

\textbf{(2) Retrieved pseudo-values.}
We also construct pseudo-values from retrieved similar products using General Multimodal Embedder (GME)~\cite{zhang2024gme}. For each sample, we retrieve the top-3 most similar products (excluding itself) that share the same target attribute and use their ground-truth values as pseudo-values. These retrieved values are often plausible but imperfect, requiring the model to verify and, when necessary, correct them using product evidence.

For each training sample, one pseudo-value is supplied while the target output remains the same ground-truth canonical value. As shown in Section~\ref{subsec:pseudo_analysis}, exposing the model to diverse pseudo-values during training is important for learning evidence-based correction.

\subsection{Self-Consistency Refinement}
\label{subsec:scr}

Although base C2P already learns to correct pseudo-values, some samples remain sensitive to the supplied pseudo-value, especially for ambiguous or confusable attributes. Different pseudo-values may still yield different predictions, indicating that the model has not yet learned stable correction behavior under pseudo-value
perturbations. To reduce this sensitivity, we introduce a second-stage training procedure called Self-Consistency Refinement (SCR).

Given a trained base C2P model, we evaluate each training sample under a predefined set of pseudo-values. If the predicted attribute value changes across pseudo-values, we mark the sample as \emph{inconsistent}. These inconsistent samples identify cases where the model's prediction is still influenced by the initial hypothesis, and therefore serve as natural targets for refinement.

We then continue training on the inconsistent subset using the same supervised AVE objective as in base training. To maintain overall stability and avoid over-specializing to the inconsistent cases, we replay a portion of the original base training data during refinement. 
SCR thus uses prediction instability under pseudo-value
perturbation as a task-specific signal for targeted refinement. As
shown in Section~\ref{subsec:scr_analysis}, this signal is enriched with hard and confusable cases, making it effective in practice.

Formally, let $\mathcal{P}(x)$ denote the predefined set of pseudo-values used to evaluate training sample $x$, and let $f_\theta(x,p)$ denote prediction of the current C2P model for sample $x$ under pseudo-value $p$. We mark $x$ as inconsistent if there exist $p_i, p_j \in \mathcal{P}(x)$ such that
\begin{equation}
    f_\theta(x, p_i) \neq f_\theta(x, p_j).
\end{equation}
Let $\mathcal{D}_{\text{inc}}$ denote the resulting inconsistent subset. SCR then trains on $\mathcal{D}_{\text{inc}}$ together with 50\% replayed base-training samples.

In our implementation, the inconsistency-mining step is performed offline during training. For each sample, we evaluate the model under a predefined set of pseudo-values, consisting of \texttt{leastV}, \texttt{invalid}, and $\mathcal{C}_{c,a}^{(\leq 20)}$, where $\mathcal{C}_{c,a}^{(\leq 20)}$ denotes the displayed option set containing up to 20 frequent canonical values for the corresponding category-attribute pair. A sample is marked as inconsistent if its prediction changes across these pseudo-values.

We apply SCR for two iterations in all main results. After base C2P training, the mined inconsistent subset is used for C2P-SCR-iter1 training. We then re-evaluate the updated model on the training set to identify the remaining inconsistent samples for C2P-SCR-iter2. In practice, the resulting inconsistent subset is relatively small: the first and second SCR iterations use roughly 5\% and 4\% of the original training set, respectively, before adding replayed samples.

\subsection{Inference with a Fixed Placeholder}
\label{subsec:inference}
Having learned correction behavior during training, C2P can use a simple fixed-placeholder design at inference time to trigger single-pass prediction without online retrieval or iterative refinement.
We therefore supply a fixed placeholder $p^\ast$ and perform a single forward pass:
\begin{equation}
    \hat{y} = f_\theta(x, p^\ast).
\end{equation}
This design keeps inference simple and efficient, which is important for industrial deployment.

In practice, different fixed placeholders are possible. Rather than relying on online candidate retrieval, C2P uses a single placeholder chosen empirically from a small set of fixed options. As shown in Section~\ref{subsec:inference_analysis}, fixed placeholders are sufficient to trigger the correction behavior learned during training, and SCR further reduces sensitivity to the specific placeholder choice. In our main setting, we use \texttt{leastV}, i.e., the least frequent canonical value for the category-attribute pair, as a practical default. We do not treat this choice as theoretically special; it is simply the best-performing fixed placeholder among those tested in our setting.

\begin{table}[t]
    \centering
    \small
    \caption{Details of two datasets.}
    \begin{tabular}{c|ccc|ccc}
    \hline
    Datasets & \#cate & \#attr & \#(cate, attr) & \#train & \#val &\#test  \\
    \hline
\textit{ImplicitAVE}	&	5	&	23	  & 25  &  54881  & 13723  &	1610    \\
\textit{AE-Product}	    &	371 &	1267  & 3457	&  233908  & 25989	 &	34042   \\
\hline
    \end{tabular}
    \label{table:datasets}
\end{table}

\section{Experiments}
\subsection{Datasets and implementation}
We evaluate C2P on both a public benchmark and a large-scale industrial dataset. \textit{ImplicitAVE}~\cite{zou2024implicitave} is an open benchmark adapted from \textit{MAVE}~\cite{yang2022mave} for implicit multimodal attribute value extraction, where the target value is not explicitly stated in the text and must be inferred from multimodal evidence. \textit{AE-Product} is a large-scale industrial dataset collected from AliExpress, covering 371 categories, 1,267 attributes, and 3,457 category-attribute pairs. Unlike \textit{ImplicitAVE}, \textit{AE-Product} also contains invalid cases in which no supported canonical value exists. Table~\ref{table:datasets} summarizes the dataset statistics. We use Micro-F1 as the primary offline metric.

We use Qwen2.5-VL-7B-Instruct~\cite{bai2025qwen25vl} as the backbone and fine-tune it with LoRA~\cite{hu2022lora} (rank 8, $\alpha=2$), implemented in PyTorch~\cite{pytorch} and optimized with AdamW~\cite{kingma2015adam}. Base C2P is trained for one epoch with learning rate $1\times10^{-4}$, weight decay $0.1$, and batch size $16$. During SCR, we continue training on the mined inconsistent subset together with 50\% replayed base-training samples to mitigate forgetting, while keeping the other settings unchanged and reducing the learning rate to $1\times10^{-5}$. All models are trained on NVIDIA A100 GPUs, while serving efficiency is measured on L20 GPUs.
 
For retrieved pseudo-values, we use the General Multimodal Embedder (GME)~\cite{zhang2024gme}, specifically \texttt{gme-Qwen2-VL-2B-Instruct}, which encodes image-title pairs into joint embeddings. We compute cosine similarity and retrieve the top-3 nearest products (excluding itself) that share the same target attribute, without applying a similarity threshold. We compare against representative baselines from four families: strong closed-source zero-shot MLLMs (GPT-4V, GPT-4o, and GPT-o1), a zero-shot multi-agent refinement method (MADIAVE), a strong supervised AVE baseline (MICE), and direct supervised fine-tuning of the same Qwen2.5-VL backbone.

\begin{table*}[t!]
    \centering
    \caption{Micro-F1 (\%) comparisons on \textit{ImplicitAVE} across categories between Correcting-to-Predict with Self-Consistency Refinement (C2P-SCR) and zero-shot/supervised fine-tuned (SFT) methods. Best results are highlighted in bold.}
    \begin{tabular}{c|ccccc|c}
        \hline
        \textbf{Methods} & \textbf{Clothing} & \textbf{Footwear} & \textbf{Jewelry} & \textbf{Food} & \textbf{Home} & \textbf{Overall} \\
        \hline
\textbf{Zero Shot Methods}	&		&		&		&		&		&		\\
GPT-4V	&	77.43	&	81.39	&	90.45	&	90.77	&	89.93	&	86.77	\\
GPT-4o	&	85.66	&	80.48	&	91.87	&	90.26	&	82.41	&	85.68	\\
GPT-o1	&	84.96	&	81.08	&	91.36	&	90.26	&	83.81	&	86.27	\\
Qwen2.5-VL-7B	&	64.16	&	61.83	&	77.72	&	74.87	&	75.71	&	71.43	\\
MADIAVE (Qwen2.5-VL-7B)	&	61.95	&	65.93	&	85.00	&	85.13	&	81.84	&	77.14	\\
\textbf{SFT Methods}	&		&		&		&		&		&		\\
MICE	&	85.40	&	83.60	&	91.82	&	\textbf{91.54}	&	\textbf{87.31}	&	87.95	\\
Qwen2.5-VL-7B	&	85.40	&	85.80	&	93.64	&	89.74	&	83.74	&	87.19	\\
C2P-SCR (ours)	&	\textbf{88.50}	&	\textbf{85.80}	&	\textbf{94.55}	&	90.26	&	87.03	&	\textbf{88.87}	\\
\hline
    \end{tabular}
    \label{table:mainresult}
\end{table*}

\subsection{Main Results}
Table~\ref{table:mainresult} reports the Micro-F1 comparisons among representative baselines on the five categories of \textit{ImplicitAVE}. Unless otherwise specified, C2P-based results in this section are reported under the default fixed-placeholder inference setting. Compared with the direct-generation baseline obtained by supervised fine-tuning the backbone, our method performs better in all categories, improving the overall score by 1.68 points and yielding a 3.29-point gain on \textit{Home}. Moreover, our method surpasses the zero-shot baseline GPT-4V and the strongest supervised fine-tuning baseline MICE by 2.10 and 0.92 points, respectively, on the full evaluation set. Notably, in \textit{Clothing} and \textit{Jewelry}, our approach achieves substantial gains of 2.84 and 2.68 points over the previous best-performing models.

Table~\ref{table:ablations} further summarizes the progressive effect of the main C2P components on both \textit{ImplicitAVE} and \textit{AE-Product}. In \#2, the placeholder \texttt{invalid} is used for both training and inference, and performance remains close to direct generation. By contrast, adding retrieved pseudo-values yields clear gains on both datasets. SCR brings further improvements on top of base C2P, reaching 88.87 on \textit{ImplicitAVE} and 91.39 on \textit{AE-Product}. These results show that the correction-based formulation is effective on both the public benchmark and the industrial setting.

\begin{table}[t]
    \centering
    \small
    \setlength{\tabcolsep}{4pt}
    \caption{Micro-F1 (\%) of progressively adding the main components of C2P on \textit{ImplicitAVE} and \textit{AE-Product}.}
    \begin{tabular}{c|ccc|cc}
    \hline
Configurations & APV & RPV & SCR & \textit{ImplicitAVE} & \textit{AE-Product} \\
    \hline
\#1 &  &  &  & 87.19 & 88.12 \\
\#2 & \checkmark &  &  & 87.00 & 88.29 \\
\#3 & \checkmark & \checkmark &  & 88.31 & 90.38 \\
\#4 & \checkmark & \checkmark & \checkmark & \textbf{88.87} & \textbf{91.39} \\
\hline
    \end{tabular}
    \label{table:ablations}
\end{table}

To better understand where the gains of C2P come from, we further partition attributes by difficulty under the direct-generation
Qwen2.5-VL SFT baseline. Specifically, we label an attribute as \texttt{hard} if the baseline achieves Micro-F1 below 0.85 on that attribute, and \texttt{easy} otherwise. Table~\ref{table:splitp} reports the results on the resulting subsets. C2P preserves strong performance on \texttt{easy} attributes while yielding substantially larger gains on \texttt{hard} ones, and C2P-SCR further amplifies these improvements. This pattern suggests that the proposed correction-based formulation is particularly helpful for hard attributes, which in practice often correspond to ambiguous or confusable cases where subtle multimodal cues are more critical for disambiguation.

\begin{table}[t]
    \centering
    \small
    \caption{Micro-F1 (\%) on easy and hard attribute subsets, with hard attributes defined by low baseline performance.}
    \begin{tabular}{c|cc|cc}
        \hline
        \multirow{2}{*}{Methods} & \multicolumn{2}{c|}{\textit{ImplicitAVE}} & \multicolumn{2}{c}{\textit{AE-Product}} \\
                &   \texttt{easy}  &  \texttt{hard}   & \texttt{easy}  & \texttt{hard} \\
        \hline
baseline	    & 91.56    & 76.89	    & 94.30	    & 66.87	    \\
C2P	            & 92.68 (+1.12)	  & 78.03 (+1.14)	 & 94.79 (+0.49)	        & 75.12 (+8.25)	        \\
C2P-SCR         & 92.72 (+1.16)    & 79.92 (+3.03)      & 95.36 (+1.06)    & 76.79 (+9.92)     \\
\hline
    \end{tabular}
    \label{table:splitp}
\end{table}

\subsection{Why Pseudo-Value Correction Helps}
\label{subsec:pseudo_analysis}
We next analyze why the correction-based formulation improves over direct generation, focusing on two training-time questions: what kind of pseudo-values are most useful during training, and how retrieved values should be used during training.

Table~\ref{table:c2p} compares different pseudo-value strategies in base C2P. To isolate the effect of pseudo-value quality from the number of training instances, both C2P-fixV and C2P-randV are matched to C2P-GME-top3 in training size. C2P-fixV, which always uses the uninformative placeholder \textit{invalid} for both training and inference, remains close to direct generation, suggesting that a trivially bypassable pseudo-value provides limited correction supervision. In contrast, C2P-randV already improves over the baseline on both datasets, showing that the correction formulation itself is beneficial even with noisy hypotheses. More importantly, retrieved pseudo-values from GME (C2P-GME-top3) perform best, indicating that plausible but imperfect hypotheses provide stronger and more informative supervision for learning correction behavior than either meaningless placeholders or arbitrary random values.

\begin{table}[t]
    \centering
    \caption{Micro-F1 (\%) of different training strategies in base Correcting-to-Predict (C2P). C2P-GME-top3 corresponds to the default base C2P model.} 
    \begin{tabular}{c|cc}
        \hline
        Methods & \textit{ImplicitAVE} & \textit{AE-Product} \\
        \hline
baseline	    &	87.19	&	88.12	\\
C2P-fixV	    &	87.13	&	88.53	\\
C2P-randV	    &	87.50	&	88.86	\\
C2P-GME-top3	&	\textbf{88.31}	&	\textbf{90.38}	\\
\hline
    \end{tabular}
    \label{table:c2p}
\end{table}

We further compare C2P with retrieval-augmented generation (RAG), which injects retrieved product cases directly into the prompt. Here the key difference is not whether retrieval is used, but how retrieved values are used. In realistic Test$\rightarrow$Train retrieval, the top-1 retrieved value matches the query ground-truth only $65.9\%$ and $55.1\%$ on \textit{ImplicitAVE} and \textit{AE-Product}, respectively, indicating that retrieved candidates are informative but imperfect. As shown in Table~\ref{table:gmetopk}, C2P-top3 consistently outperforms RAG-top3 on both datasets while using substantially fewer input tokens. This suggests that, for AVE, retrieved values are more effective when used as compact hypotheses to be verified and corrected than when used as prompt context. Since top-3 provides the best trade-off in our setting, we use it as the default configuration in base C2P.

\begin{table}[t]
    \centering
    \caption{Micro-F1 (\%) and inference-time efficiency metrics (average input tokens and queries per second, QPS) comparing the base Correcting-to-Predict (C2P) with Retrieval-Augmented Generation (RAG) under top-k retrieval.}
    \begin{tabular}{c|cc|cc}
        \hline
        Methods & \textit{ImplicitAVE} & \textit{AE-Product} & avg. tokens & QPS \\
        \hline
baseline	&	87.19	&	88.12    &	201   & 5.18 \\
\hline                         
RAG-top1	&	87.71	&	89.03    &	340   & 2.69 \\
RAG-top3	&	87.77	&	89.13    &	585   & 1.45 \\
RAG-top5	&	87.77	&	89.26    &	832   & 1.01 \\
\hline                        
C2P-top1	&	87.87	&	89.56    &	233   & 4.88 \\
C2P-top3	&	\textbf{88.31}	&	\textbf{90.38}    &	233  & 4.77 \\
C2P-top5	&	88.00	&	89.92    &	234   & 4.92 \\
\hline
    \end{tabular}
    \label{table:gmetopk}
\end{table}

\subsection{Self-Consistency Refinement Analysis}
\label{subsec:scr_analysis}

We next examine why Self-Consistency Refinement (SCR) is effective.
The starting point is that, even after base C2P training, some samples remain sensitive to the supplied pseudo-value. SCR targets this residual instability by mining samples whose predictions change
under pseudo-value perturbations and retraining them.

A natural question is whether this inconsistency signal identifies
meaningful error-prone cases or merely reflects random variation.
Empirically, the mined subset is strongly enriched with difficult attributes: the proportion of hard attributes increases from 28.29\%
to 68.14\% on \textit{ImplicitAVE} and from 33.71\% to 58.58\% on \textit{AE-Product}. This suggests that prediction inconsistency is a useful task-specific signal for locating ambiguous and confusable samples where stable evidence-based correction behavior has not yet been learned.

We next verify that SCR improves the intended failure modes rather
than acting as generic extra fine-tuning. To this end, we measure
three diagnostics under fixed-placeholder inference: PRR, the fraction of predictions identical to the supplied pseudo-value; E-PRR, the same ratio restricted to error cases; and E-FVR, the fraction of error cases predicted as the most frequent value for the corresponding category-attribute pair. As shown in Table~\ref{table:scr_error_metrics}, after SCR, PRR changes only marginally on ImplicitAVE (14.05$\rightarrow$13.93), while E-PRR and E-FVR drop from 14.97/24.06 to 13.97/21.79.
This suggests that SCR improves accuracy not by indiscriminately
suppressing retention behavior, but by reducing errors that remain
overly driven by the supplied pseudo-value or by frequent-value priors. We observe the same directional trend on AE-Product.

\begin{table}
    \centering
    \small
    \caption{Pseudo-value-sensitive error metrics before and after SCR on \textit{ImplicitAVE} and \textit{AE-Product}.}
    \begin{tabular}{c|ccc|ccc}
        \hline
        \multirow{2}{*}{Method} & \multicolumn{3}{c|}{\textit{ImplicitAVE}} & \multicolumn{3}{c}{\textit{AE-Product}} \\
         & PRR & E-PRR & E-FVR & PRR & E-PRR & E-FVR \\
        \hline
        C2P & 14.05 & 14.97 & 24.06 & 10.14 & 47.99 & 72.41 \\
        C2P-SCR & 13.93 & 13.97 & 21.79 & 8.87 & 45.57 & 71.64 \\
        \hline
    \end{tabular}
    \label{table:scr_error_metrics}
\end{table}

Table~\ref{table:scr_variant_f1} further compares SCR with a curriculum-learning-style baseline (C2P-CL) and multiple refinement iterations under the default \texttt{leastV} inference setting. Unlike SCR, C2P-CL selects samples based on prediction difficulty rather than pseudo-value stability. C2P-CL remains slightly below base C2P, whereas SCR yields clear gains on both datasets. Moreover, most of the gain is already achieved in the first refinement iteration, with the second iteration providing only marginal improvement.

\begin{table}
    \centering
    \small
    \caption{Micro-F1 (\%) of refinement variants.}
    \begin{tabular}{l|cc}
        \hline
        Method & \textit{ImplicitAVE} & \textit{AE-Product} \\
        \hline
        C2P & 88.31 & 90.38 \\
        C2P-CL & 88.06 & 90.33 \\
        C2P-SCR-iter1 & 88.87 & 91.36 \\
        C2P-SCR-iter2 & 88.87 & 91.39 \\
        \hline
    \end{tabular}
    \label{table:scr_variant_f1}
\end{table}

Overall, these results show that SCR acts as a targeted refinement stage for unstable correction behavior rather than as a generic second-stage fine-tuning procedure.

\subsection{Fixed-Placeholder Inference Analysis}
\label{subsec:inference_analysis}

We next analyze the inference-time design of C2P. A key question is whether the learned correction behavior still requires retrieval-time pseudo-values, or whether it can be triggered by a fixed placeholder in a single forward pass. Table~\ref{table:placeholder_f1} reports the Micro-F1 under different inference-time pseudo-values, including fixed placeholders (\texttt{invalid}, \texttt{mostV}, and \texttt{leastV}) and GME-top3, where one retrieved pseudo-value is sampled from the top-3 retrieved candidates.

\begin{table}
    \centering
    \small
    \caption{Micro-F1 (\%) under fixed and retrieved pseudo-values at inference time.}
    \begin{tabular}{c|cc|cc}
        \hline
        \multirow{2}{*}{Placeholder} & \multicolumn{2}{c|}{C2P} & \multicolumn{2}{c}{C2P-SCR} \\
         & \textit{ImplicitAVE} & \textit{AE-Product} & \textit{ImplicitAVE} & \textit{AE-Product} \\
        \hline
        \texttt{invalid} & 87.94 & 89.73 & 88.87 & 91.26 \\
        \texttt{mostV} & 87.50 & 89.53 & 88.74 & 91.30 \\
        \texttt{leastV} & 88.31 & 90.38 & 88.87 & 91.39 \\
        \texttt{GME-top3} & 87.75 & 89.75 & 88.81 & 91.35 \\
        \hline
    \end{tabular}
    \label{table:placeholder_f1}
\end{table}

For the base C2P model, \texttt{leastV} yields the best performance on both \textit{ImplicitAVE} and \textit{AE-Product}, achieving 88.31 and 90.38, respectively. However, the more important observation is that inference-time retrieval is not necessary to realize the benefit of C2P: fixed placeholders already trigger effective correction behavior, and in our setting perform comparably to or better than sampled retrieved pseudo-values. In particular, retrieval-based GME-top3 reaches 87.75 and 89.75, while \texttt{invalid} gives 87.94 and 89.73.

After SCR, the gap across inference-time pseudo-values becomes even smaller. On \textit{ImplicitAVE}, C2P-SCR achieves 88.87 with both \texttt{invalid} and \texttt{leastV}, compared with 88.81 under GME-top3; on \textit{AE-Product}, the corresponding numbers are 91.26, 91.39, and 91.35. This indicates that SCR further reduces sensitivity to the specific pseudo-value supplied at inference time.

We therefore adopt \texttt{leastV} as the default inference placeholder, but emphasize that C2P does not rely on retrieval-time pseudo-values for its gains. Overall, these results support a simple and practical deployment design: C2P can perform retrieval-free, single-pass inference with a fixed placeholder, while maintaining accuracy comparable to or better than inference with retrieved pseudo-values.

\subsection{Efficiency and Online Deployment}

A practical advantage of C2P is that correction behavior is learned during training, while inference remains retrieval-free and single-pass. Table~\ref{table:efficiency_deploy} reports the serving efficiency. Compared with direct generation and RAG-top3, C2P preserves near-baseline serving throughput (4.77 vs. 5.18 QPS on a single L20 GPU) while avoiding the substantial latency increase of retrieval-based inference (1.45 QPS for RAG-top3).

\label{subsec:deployment}
\begin{table}
    \centering
    \small
    \caption{Inference efficiency in industrial deployment.}
    \begin{tabular}{l|cc}
        \hline
        Method & \begin{tabular}[c]{@{}c@{}}QPS\\(single L20)\end{tabular} & \begin{tabular}[c]{@{}c@{}}Time for 1M samples\\(8 L20 GPUs, hours)\end{tabular} \\
        \hline
        Direct Generation & 5.18 & 6.7 \\
        C2P & 4.77 & 7.3 \\
        RAG-top3 & 1.45 & 23.9 \\
        \hline
    \end{tabular}
    \label{table:efficiency_deploy}
\end{table}

The C2P framework has been deployed across multiple production scenarios on AliExpress. These deployment gains are consistent with
two effects observed offline: improved extraction quality and an expanded set of category-attribute pairs that satisfy deployment-time
accuracy requirements. In a 7-day online A/B test with 10\% of live traffic, C2P-SCR improved business metrics over the production Qwen2.5-VL SFT baseline across listing, search, and product-card scenarios: deployable category-attribute pairs increased by \textbf{22.6\%}; the GMV share of products with complete core attributes improved by \textbf{9.8\%}; seller adoption in listing rose from \textbf{83\%} to \textbf{85\%}; search obtained \textbf{37.2\%} more relevant filter options, with \textbf{+15.6\%} filter usage and \textbf{+3.3\%} CTR in filtered sessions; and product-card highlights improved overall CTR by \textbf{+0.94\%} and buyers by \textbf{+0.54\%}.

\section{Conclusion}
\label{sec:conclusion}

This paper presented \textbf{Correcting to Predict (C2P)}, a practical framework for multimodal attribute value extraction (AVE) in e-commerce. C2P reformulates AVE as correcting an initial pseudo-value with multimodal evidence, rather than directly generating an attribute value from scratch. By training with diverse pseudo-values, the model learns when a supplied hypothesis should be retained or revised, and a subsequent Self-Consistency Refinement (SCR) stage further reduces sensitivity to pseudo-value perturbations.

Experiments on the public \textit{ImplicitAVE} benchmark, a large-scale industrial dataset, and online A/B tests on AliExpress show that C2P consistently outperforms strong baselines, especially on hard attributes that are often ambiguous or confusable, while supporting efficient single-pass deployment without online retrieval or iterative refinement. We hope this work encourages further study of correction-based formulations for multimodal prediction tasks in real-world systems.

\section*{GenAI Usage Disclosure}

We acknowledge the use of Generative AI (GenAI) tools in the preparation of this work. GenAI tools were only used to improve the spelling and grammar of certain passages. All uses of GenAI tools complied with the ACM Authorship Policy on GenAI usage.

\bibliographystyle{ACM-Reference-Format}
\balance
\bibliography{reference}


\end{document}